\documentclass[prd,twocolumn,superscriptaddress,showpacs,nofootinbib,preprintnumbers]{revtex4}

\usepackage{amsmath}
\usepackage{amsfonts}
\usepackage{graphicx}
\usepackage{dcolumn}

\def\be{\begin{equation}}
\def\ee{\end{equation}}
\def\ba{\begin{eqnarray}}
\def\ea{\end{eqnarray}}
\def\bs{\begin{subequations}}
\def\es{\end{subequations}}
\def\tX{\tilde{X}}
\def\tv{\tilde{v}}
\def\tT{\tilde{T}}

\newcommand{\rd}{{\rm d}}

\usepackage{color}

\begin{document}

\title{Inflation from D3-brane motion in the background of D5-branes}

\bigskip

\author{Sudhakar Panda}
\affiliation{Harish-Chandra Research Institute, Chhatnag Road,
Jhusi, Allahabad-211019, India}
\email{panda@mri.ernet.in}

\author{M.~Sami}
\affiliation{Department of Physics, Jamia Millia,
New Delhi-110025, India}
\email{sami@jamia-physics.net}

\author{Shinji Tsujikawa}
\affiliation{Department of Physics, Gunma National College of
Technology, Gunma 371-8530, Japan}
\email{shinji@nat.gunma-ct.ac.jp}

\author{John Ward}
\affiliation{Department of Physics, Queen Mary, University of
London, Mile End Road, London, E1 4NS U.K.}
\email{j.ward@qmul.ac.uk}

\date{\today}

\preprint{KEK-TH-1064, QMUL-PH-05-21}

\begin{abstract}

We study inflation arising from the motion of a Bogomol'nyi-Prasad-Sommerfield
(BPS) D3-brane in
the background of a stack of $k$ parallel D5-branes. There are
two scalar fields in this set up-- (i) the radion field $R$, a
real scalar field, and (ii) a complex tachyonic scalar field
$\chi$ living on the world volume of the open string stretched
between the D3 and D5 branes. We find that inflation is
realized by the potential of the radion field, which satisfies
observational constraints coming from the Cosmic Microwave
Background. After the radion becomes of order the string length
scale $l_s$, the dynamics is governed by the potential of the
complex scalar field. Since this field has a standard kinematic
term, reheating can be successfully realized by the mechanism of
tachyonic preheating with spontaneous symmetry breaking. 

\end{abstract}

\maketitle

\section{Introduction}

There has been a resurgence of interest in the time-dependent
dynamics of extended objects found in the spectrum of string
theory, inspired in part by Sen's construction of a boundary state
description of open string tachyon condensation.
See, for example, Ref.~\cite{senrev} for review.
This description has been
supplemented well by an effective theory described by a
Dirac Born Infeld (DBI) type
action for the tachyon field \cite{DBI}. More recent work has
focused on the dynamics of a probe BPS D-brane in a variety of
gravitational backgrounds inspired by the observation that there
exists a similarity between the late time dynamics of the probe
D-branes and the condensation of the open string tachyon on the
world-volume of non-BPS brane in flat space. The latter dynamics is
also described by the DBI action \cite{kutasov}, see also
Refs.~\cite{Sahakyan,pani,TW}. Both systems describe rolling matter fields which
have a vanishing pressure at late times. As a result we can,
through an appropriate field transformation, investigate the
physics of gravitational backgrounds in terms of non-trivial
fields on a brane in flat space using the DBI effective action.
This has led to the interesting proposal that the open string tachyon
may be geometrical in nature.

Many of the backgrounds that have been probed in this manner have
been supergravity (SUGRA) brane solutions of type II string theory. By
ensuring that the number of background branes is large we can
trust our SUGRA solutions. Moreover we can neglect any back
reaction of the probe upon the background geometry. This allows us
to use the DBI action to effectively determine the relativistic
motion of extended objects in a given background. Quantum
corrections can also be calculated in those backgrounds that have
an exact Conformal Field Theory (CFT) description \cite{Sahakyan}. 
The dynamics of branes in various
backgrounds is expected to be relevant for string theory inspired
cosmology, just as in the case of open string tachyon matter
\cite{tachinfl} since the field (radion) which parameterizes  the
distance between the probe brane and the static background branes
is a scalar and may be a potential candidate for being the inflaton.

One of the most important theoretical advances in modern cosmology
has been the inflationary paradigm, which relies on a scalar field
to solve the horizon and flatness problems in the early universe
(see Refs.~\cite{inflation} for review).
Recent observations from WMAP \cite{WMAP}, SDSS \cite{SDSS} and
2dF \cite{2dF} impose tight restrictions on the possible
mechanisms that can satisfy the
paradigm \cite{general}, and hence provide the interesting possibility for
us to test string theoretic inflation models.
The observations of Supernova Ia \cite{SNIa} also
suggest that our universe is currently undergoing a period of
accelerated expansion, which is attributed to dark energy.
It still remains a fundamental problem to describe dark energy
in a purely stringy context, although
there has been several recent developments \cite{darkenergy}.

There have been many attempts to embed inflation within string
theory. The most popular approach has been to invoke the use of
the open string tachyon living on a non-BPS brane as a candidate
for the inflaton \cite{tachinfl} (see Refs.~\cite{cospep} for a
number of cosmological aspects of tachyon). Unfortunately it has
been shown that this cannot be implemented in a consistent manner,
at least in the simplest scenarios \cite{KL02,Frolov}. The other
common approach is so-called D-brane inflation in which the
separation between branes plays the role of the inflation
\cite{Dvali,braneantibrane,Garcia,Hirano}. In particular this is
well accommodated in a form of hybrid inflation where tachyonic
open string fluctuations are the fields which end inflation, and
another field is chosen to be the inflaton. These open string
fluctuations arise in the context of all D-brane cosmological
models once the branes are within a string length of one another.
A concrete example of this occurs in brane/anti-brane inflation
\cite{braneantibrane} and recently in the context of more
phenomenological warped compactifications \cite{kklmmt} (see also
Refs.~\cite{dbranepapers}). It should be noted that most of the
work done in this direction assumes that the dimensionalities of
the brane and anti-brane are the same apart from D3/D7 brane
inflation models studied in Refs.~\cite{D3D7} which does not
include the open string tachyon dynamics at late times. On the
contrary, in our model a probe D3 brane is used to lead to
inflation in the presence of static D5 branes (see
Refs.~\cite{TWinf,PST} for related works) and the open string
tachyon dynamics naturally comes in. In any event there has been
very little work done on trying to understand the relationship
between inflation and the current dark energy phase which we
observe.

A potential solution for both inflation and dark energy, in this
context, can be obtained as a mixture of these two scenarios. We
require a mechanism which drives inflation independently of the
open string tachyon, but then falls into the tachyonic state at
late times. This can be achieved by considering the motion of a
D3-brane in a type IIB background. By switching to our
holographic picture of a non-trivial field on a non-BPS brane
\cite{kutasov} we will find that the radion field naturally exits
from inflation once it reaches a critical velocity. If this occurs
at a distance larger than the string length, we can then use the
open string tachyon, which sets in at a distance equal to or less
than the string length, to explain the dark energy content of the
universe.

In this paper we aim to explore the motion of a probe D3-brane in
the background of $k$ coincident, static D5-branes. For simplicity
we will neglect any closed string radiation which would be emitted
from the probe brane as it travels down the throat generated by
the background branes. We will also neglect any gauge fields which
may exist on the D3-brane world-volume.
Note that this is S-dual to the solution considered in
Ref.~\cite{kutasov}. In order to make contact with
four dimensional physics we must consider the dual
picture of a non-trivial field on a non-BPS brane in flat space,
where we also toroidally compactify the remaining six
dimensions\footnote{This is not necessary if we consider holographic
cosmology as in Ref.~\cite{mirage}.}. We will assume that there is some
mechanism which freezes the various moduli of the compactification
manifold so that they do not appear in the effective action. The
resulting theory should represent the leading order contribution
which would arise from compactifying the full type IIB background.
At distances large compared to the string scale, the DBI
description is known to be valid, however once the probe brane
approaches small distances (order of string length scale) we must
switch to the open string analysis. Open strings will stretch from
the D3-brane to the D5-branes, and their fluctuation spectrum
contains a tachyonic mode. Thus when the separation is order of
the string length, the DBI description will no longer be valid and
we must resort to a purely open string analysis. We expect that
inflation will occur in the large field (radion) regime and it
ends before the separation comes closer to string length and that
as the branes get closer, the open string tachyon reheats
the universe.

This paper is organized as follows. In the next section we describe
the dynamics of a single probe D3-brane in the presence of a large
number of static background D5-branes. Because of the
dimensionalities of the branes we expect to find an open string
tachyonic mode once we begin to probe distances approaching the
string length \cite{branonium}. In section III, we present
the inflation dynamics and
observational constraints on the various parameters of our model.
In section IV, we discuss the role of
the open string tachyon after the inflationary phase and the
possibility of reheating in our model and a brief discussion on
dark energy. In the last section, we present some of our
conclusions and future outlook.

\section{D3-brane dynamics in D5-brane background}

In this section we analyze the motion of a probe BPS D3-brane in
the background generated by a stack of coincident and static BPS
D5-branes. The background fields, namely the metric, the dilaton
$(\phi)$ and the Ramond-Ramond (RR) field $(C)$
for a system of $k$ coincident
D5-branes are given by \cite{Callan,kutasov}
\begin{eqnarray}
g_{\alpha \beta} &=& F^{-1/2} \eta_{\alpha \beta},~~
g_{mn} =F^{1/2} \delta_{mn},
\nonumber\\
e^{2\phi} &=& F^{-1} = ~ C_{0...5},
   ~~F = 1 +\frac{kg_sl_s^2}{r^2}\,,
\label{bfields}
\end{eqnarray}
where $\alpha,\beta = 0,..,5; ~ m,n = 6,...,9$ denote the indices
for the world volume and the transverse directions respectively
and $F$ is the harmonic function describing the position of the
$k$ D5-branes and satisfying the Green function equation in the
transverse four dimensional space.
Here $g_s$ and $l_s = \sqrt{\alpha'} $ are the string coupling and
the string length, respectively.
$r$ is the radial coordinate away from the D5-branes
in the transverse direction.
The solution parameterizes a throat-like geometry which
becomes weakly coupled as we approach the source branes.

The motion of the D3-brane in the above background can be studied in
terms of an effective DBI action, on its world volume,
given by \cite{kutasov}
\begin{equation}
{\mathcal S_0} = -\tau_3\int {\rm d}^4x
F^{-1/2} \sqrt{1 + F\partial_{\alpha}
R \partial^{\alpha} R}\,,
\label{dbi}
\end{equation}
where $\tau_{3}$ is the tension of the 3-brane.
Here the motion of the probe brane is restricted to be purely radial
fluctuation, denoted by the mode $R$, along the common four
dimensional transverse space. This action is the same as that
considered in Ref.~\cite{kutasov}. The background considered here is
the S-dual to the background considered there and we have not kept
the contribution of the RR fields in the action.
The form of the above action resembles the DBI action
of the tachyon field in the open string ending
on a non-BPS D3-brane in a flat background.
This is given by
\begin{equation}
{\mathcal S_1}=- \int {\rm d}^4x
V(T) \sqrt{1 + \partial_{\alpha} T
\partial^{\alpha}T}\,.
\label{tach}
\end{equation}

Comparison of the above two actions defines
a ``tachyon'' field $T$ by the relation:
\begin{equation}
\frac{{\rm d}T}{{\rm d}R} = \sqrt{F (R)} =
\sqrt{1+L^2/R^2}\,,
\label{tachyon}
\end{equation}
where
\begin{equation}
L \equiv \sqrt{kg_{s}}l_s\,.
\end{equation}
In terms of this field the ``tachyon potential''
in Eq.~(\ref{tach}) is given by
\begin{equation}
\label{tachpo}
V=\frac{\tau_3}{\sqrt{F(R)}}=
\frac{\tau_3}{\sqrt{1+L^2/R^2}}\,.
\end{equation}
One can solve Eq.~(\ref{tachyon}) for the $T(R)$ and
find it to be a monotonically increasing function \cite{kutasov}:
\begin{equation}
T(R) = \sqrt{L^2 + R^2} + {1\over 2}L \ln {{\sqrt{L^2 + R^2} - L}
\over {{\sqrt{L^2 + R^2} + L}}}.
\end{equation}

This function is non-invertible but can be simplified by exploring limits of
the field space solution.
As $R\rightarrow 0$ we have $T(R) \rightarrow  -\infty$
with dependence
\begin{equation}
T(R\rightarrow 0) \simeq L\ln {{R\over L}}\,.
\end{equation}
As $R\rightarrow \infty$ we have $T(R) \rightarrow \infty$
with
\begin{equation}
T(R\rightarrow \infty) \simeq R\,.
\end{equation}
The effective potential in these two asymptotic regions
is given by:
\begin{eqnarray}
\label{pot2}
\frac{V(T)}{\tau_3} &\simeq & \exp
\left(\frac{T}{L}\right)
\qquad {\rm for} \quad T\rightarrow -\infty \,,\\
&\simeq & 1- \frac{1}{2}\frac{L^2}{T^2}
\qquad {\rm for} \quad T \rightarrow \infty \,.
\label{pot}
\end{eqnarray}
%

\begin{figure}
\includegraphics[height=2.8in,width=3.4in]{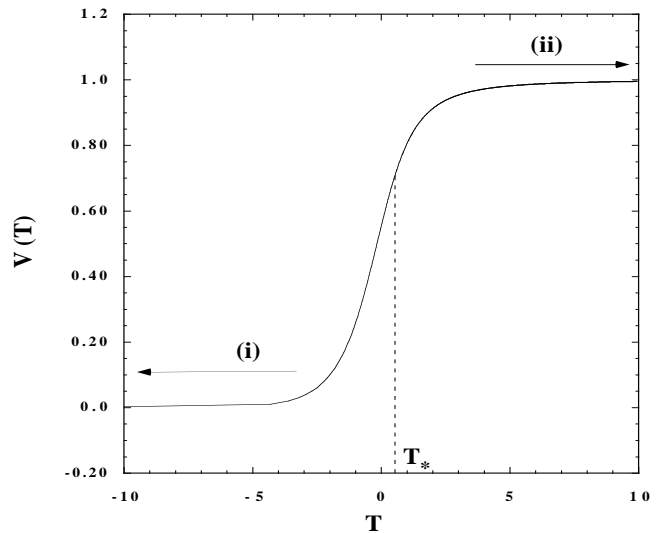}
\caption{The potential of the field $T$. The value
$T_{*}=[\sqrt{2}+{\rm ln}\,(\sqrt{2}-1)]L$ is determined by the
condition $R=L$. The potential of the region (i) is
approximately given by $V(T)=\tau_{3} \exp (T/L)$, whereas
$V(T)=\tau_{3} (1-L^2/2T^2)$ in the region (ii). }
\label{potential}
\end{figure}

Thus in the limit $T\rightarrow -\infty$, corresponding to
$R\rightarrow 0$, one observes that the potential goes to zero
exponentially (see Fig.~1). This is consistent with the late time behavior
for the open string tachyon potential in the rolling tachyon
solutions and leads to exponential decrease of the pressure at
late times \cite{sen2}. The ``tachyon field" has a geometric
meaning signifying the distance between the probe brane and the
D5-branes. At large distances, the DBI action interpolates
smoothly between standard gravitational attraction among the probe
and the background branes and a ``radion matter" phase when the
probe brane is close to the five branes. The transition between
the two behaviors occurs at $R \sim L$.

It is important to note that when the probe brane is
within the distance $R\sim l_s$, the
above description in terms of the closed string background is
inappropriate and the system should be studied using upon strings
stretched between the probe brane and the five branes. To be more
precise, when the probe brane comes to within a distance between
$l_s$ from the D5-branes, a tachyon appears in the open string
spectrum and in principle the dynamics of the system will be
governed by its condensation from that point on.

Thus the full dynamics can be divided into two regimes.
When the distance $R$ between the D3-brane and the D5-branes
is much smaller than $L$
but larger than $l_s$, we can describe the dynamics of the radial
mode $R(x^{\mu})$ by the tachyon matter Lagrangian (\ref{tach})
with an exponentially decaying potential given by (\ref{pot2})
(note that $T$ is going toward $-\infty$). On the contrary,
when $R$ is of the order of $l_s$, the dynamics would be be governed
by the conventional Lagrangian describing the complex tachyonic
scalar field $\chi$ present in the open string stretched between
the D3-brane and the $k$ D5-branes. The potential for such open
string tachyon field has already been calculated \cite{Gava}. Thus
the dynamics of $\chi$ is described by the action:
\begin{equation}
{\mathcal S_2} = \int {\rm d}^4x [-\partial_{\alpha} \chi
\partial^{\alpha} \chi^* -U(\chi, \chi^*)]\,,
\label{chi}
\end{equation}
where the potential, up to quartic order, is
given by:
\begin{equation}
U (\chi, \chi^*) =
\frac{1}{4\pi^4 l_s^4 g_s k}
\left[ \pi (k+1) (\chi \chi^*)^2-v \chi \chi^* \right]\,.
\label{chipot}
\end{equation}
Note that $\chi$ and $v$ are dimensionless quantities.
Here $v$ is a small parameter ($v \ll k$)
corresponding to the volume of a two torus. This arises as we are
toroidally compactifying the directions transverse to the
D3-brane, but parallel to the D5-branes, in order to describe the
dynamics of the open string tachyon. When we map the theory to our
purely $3+1$ dimensional subspace, we will neglect any string
winding modes arising from this torus. Furthermore it can be seen
that our fully compactified theory is actually not $T^6$ but the
product space $T^4 \times T^2$ but for simplicity we shall
assume that the relevant radii are approximately equal.

Let us briefly recapitulate and consider the bulk dynamics in more
detail. At distances larger than the string length we know that
the DBI action provides a good description of the low energy
physics for a probe brane in the background geometry. As mentioned
in the introduction, the D3-brane is much lighter than the
coincident D5-branes and so we can neglect the back reaction upon
the geometry. Furthermore the SUGRA solution indicates that the
string coupling tends to be zero as we probe smaller distances,
providing a suitable background for perturbative string theory and
implying that we can trust our description down to small distances
without requiring a bound on the energy \cite{kutasov}.

Because of the dimensionalities of the branes in the problem there is no
coupling of the D3-brane to the bulk RR six form. This
is because the only possible Wess-Zumino interaction between the
probe brane and the background can be through the self dual field
strength $\tilde f = {\rm d}\tilde C_{(4)}$. However this field
strength must be the Hodge dual of the background field strength -
which is given here by $f = {\rm d} C_{(6)}$ for D5-branes - clearly
this inconsistency implies that the coupling term will vanish. For
a more detailed explanation of the more general case we refer the
reader to the paper \cite{branonium}, however the basic result for our
purpose is that there is only a non-zero interaction term when
either the dimensionality of probe and background branes are the
same, or they add up to six. The probe brane however does possess
its own RR charge which ought to be radiated as the
brane rolls in the background, but for simplicity we will neglect
this in our analysis.

The energy-momentum tensor density of the probe brane in the background can
be calculated as
\begin{eqnarray}
T_{ab} &=& \frac{\tau_3}{\sqrt{F}} \Biggl(\frac{F \partial_a R \partial_b
R}{\sqrt{1+ F \eta^{cd} \partial_c R \partial_d R}} \nonumber  \\
& &- \eta_{ab} \sqrt{1+F \eta^{cd} \partial_c R \partial_d R} \Biggr),
\end{eqnarray}
where the roman indices are directions on the world-volume.
As we are only interested in homogenous scalar fields in this paper,
we find that this expression reduces to
\begin{eqnarray}
T_{00} &=& \frac{\tau_3}{\sqrt{F}\sqrt{1-F\dot{R}^2}}\,, \nonumber \\
T_{ij} &=& -\frac{\tau_3 \delta_{ij}\sqrt{1-F\dot{R}^2}}{\sqrt{F}},
\end{eqnarray}
where $i, j$ are now the spatial directions on the D3-brane.

Using the energy conservation we can obtain the equation of motion
for the probe brane in our background and estimate its velocity.
By imposing the initial condition that the velocity is zero at the
point $R = R_0$ we find that the expression for the velocity
reduces to
\begin{equation}
\dot{R}^2 = \frac{R^2 L^2}{(R^2+ L^2)^2}
\left(1-\frac{R^2}{R_0^2} \right),
\end{equation}
which is obviously valid for $R \le R_{0}$ and in fact as expected
it vanishes identically at $R=R_0$. We typically would expect
$R_0$ to be extremely large. Note that in the two asymptotic
regions of small and large $R$ the velocity is tending to zero.
This is understood because the throat geometry acts as a
gravitational red-shift, giving rise to D-cceleration phenomenon
\cite{tong}. It should be emphasised that the asymptotic limit $R \to 0$ is
unphysical
because the DBI is not valid once we reach energies of the order
of string mass $M_s$,
and so it is not strictly correct to say that the velocity goes to
zero in the small $R$ approximation.
However note that when $R \rightarrow l_s$
we have $\dot{R}^2 \sim l_s^2/L^2 = 1/kg_s$ which
is also negligibly small for large $k$. From our perspective this
implies that the kinetic energy of the scalar field become
sub-dominant at small distances.
It is essentially frozen out and the dynamics of the open
string tachyonic modes come to dominate.
Once the probe brane reaches distances comparable with the string
length our closed string description is no longer valid. Instead
we must switch over to an open string description of the tachyonic
modes $\chi$ described by the action (\ref{chi}).

It is worth pointing out that our discussion so far seems to
suggest that the radionic mode and the open string tachyonic mode
which are being described by two different action functionals have
nothing in common and can be described independent of each other.
However, it is not so. First the number of background branes have
to be same. Secondly, unlike the open string tachyon on the world
volume of a non-BPS brane or a brane/anti-brane pair, the dynamics
of the tachyon on the open string connecting a BPS D$p$-brane and a
BPS D$(p+2)$-brane is not described by a DBI type action. If this
would have been the case, the above two fields could have been
combined together with keeping in mind about their region of
validity.

However, even in the present context we can combine the
two actions by introducing an interaction term like $\lambda T^2\chi^2$
where the coupling $\lambda$ will be zero for values of
the field $T$ corresponding to $R$ greater than $l_s$.
Provided that inflation ends for $R>l_s$, this term
does not affect the dynamics of inflation and
for simplicity we have ignored it in the action functional.
However, such a term may play an important role
in a possible reheating phase. We can now
proceed with our analysis of inflation using the full form of the
harmonic function - which specifies the scalar field potential in terms of
the geometrical tachyon field rather than the radion field.

\section{Inflation and observational constraints from CMB}

In this section we shall discuss the dynamics of inflation and
observational constraints on the model (\ref{tachpo}) from
CMB. Introducing a dimensionless quantity $x \equiv R/L$, the full
potential (\ref{tachpo}) of the field $T$ is written as
\begin{equation}
\label{pox}
V=\frac{x}{\sqrt{x^2+1}}\tau_{3}\,,
\end{equation}
where $\tT \equiv T/L$ is related to $x$ via
\begin{equation}
\frac{\rd \tT}{\rd x}=\frac{\sqrt{x^2+1}}{x}
=\frac{1}{\tilde{V}}\,,
\end{equation}
where $\tilde{V} \equiv V/\tau_{3}$.
We require that $R$ is larger than
$l_{s}$, which translates into the condition
$x>1/\sqrt{kg_s}$.

In a flat Friedmann-Robertson-Walker background
with a scale factor $a$ the field equations are \cite{cospep}
\ba
\label{tach1}
& & H^2
=\frac{1}{3M_p^2}
\frac{V(T)}{\sqrt{1-\dot{T}^2}}\,, \\
\label{tach2}
& & \frac{\ddot{T}}{1-\dot{T}^2}+3H\dot{T}
+\frac{V_T}{V}=0\,,
\ea
where $H \equiv \dot{a}/a$ is the Hubble rate,
$V_{T} \equiv {\rm d}V/{\rm d}T$,
and $M_{p}=1/\sqrt{8\pi G}$ is the 4-dimensional
reduced Planck mass
($G$ is the gravitational constant).

Combining Eq.~(\ref{tach1}) and (\ref{tach2})
gives the relation $\dot{H}/H^2=-3\dot{T}^2/2$.
Then the slow-roll parameter is given by
\begin{eqnarray}
\label{ep}
\epsilon  &\equiv& -\frac{\dot{H}}{H^2}=\frac32 \dot{T}^2
\simeq \frac{M_{p}^2}{2}
\frac{V_{T}^2}{V^3} \nonumber \\
&=& \frac{1}{2s} \frac{\tilde{V}_x^2}
{\tilde{V}} = \frac{1}{2s} \frac{1}
{x(x^2+1)^{5/2}}\,,
\end{eqnarray}
where
$s$ is defined by
\begin{equation}
\label{sdef}
s \equiv \frac{L^2 \tau_{3}}{M_{p}^2}\,.
\end{equation}
In deriving the slow-roll parameter we used the
slow-roll approximation $\dot{T}^2 \ll 1$ and
$|\ddot{T}| \ll 3H |\dot{T}|$
in Eqs.~(\ref{tach1}) and (\ref{tach2}).
Equation (\ref{ep}) shows that $\epsilon$ is a decreasing function
in terms of $x$. Hence $\epsilon$  increases as the field evolves
from the large $R$ region to the small $R$ region,
marking the end of inflation at $\epsilon=1$.

The number of $e$-foldings from the end of inflation is
\begin{eqnarray}
N &\equiv& \int_{t}^{t_f} H {\rm d}t
\simeq \int^T_{T_{f}} \frac{V^2}{M_{p}^2V_T}
{\rm d}T \nonumber \\
&=&
s \int^x_{x_{f}} (x^2+1)^{3/2} {\rm d}x\,.
\end{eqnarray}
This is integrated to give
\begin{equation}
\label{Nfx}
N=s[f(x)-f(x_{f})]\,,
\end{equation}
where
\begin{eqnarray}
f(x) &=& \frac14 x(x^2+1)^{3/2}+\frac38 x \sqrt{x^2+1}
\nonumber \\
& & +\frac38 {\rm ln} \left| x+\sqrt{x^2+1} \right|\,.
\end{eqnarray}
The function $f(x)$ grows monotonically from
$f(0)=0$ to $f(\infty)=\infty$ with the increase of $x$.
In principle we can obtain a sufficient amount of inflation
to satisfy $N>70$ if either $s$ or $x$ is large.

In order to confront with observations we need to
consider the spectra of scalar and tensor perturbations
generated in our model.
The power spectrum of scalar metric perturbations is
given by \cite{Hwang,SV04,GST}
\begin{eqnarray}
{\cal P}_{\rm S} &=&
\frac{1}{12\pi^2 M_{p}^6}
\left( \frac{V^2}{V_T} \right)^2
=\frac{\tau_{3}^2L^2}{12\pi^2M_{p}^6}
\left(\frac{\tilde{V}}{\tilde{V}_x}\right)^2 \nonumber \\
&=& \frac{s^2}{12\pi^2 kg_{s} (l_{s}M_p)^2}
x^2(x^2+1)^2 \,.
\label{powersca}
\end{eqnarray}
The COBE normalization corresponds to ${\cal P}_{\rm S}=2 \times
10^{-9}$ around $N=60$ \cite{inflation}, which gives
\begin{eqnarray}
\label{COBE}
kg_{s} (l_{s} M_{p})^2=\frac{10^9}{24\pi^2}
s^2x_{60}^2(x_{60}^2+1)^2\,.
\end{eqnarray}
The spectral index of curvature perturbations is
given by \cite{Hwang,SV04,GST}
\begin{eqnarray}
n_{\rm S}-1 &=& -4\frac{M_p^2V_{T}^2}{V^3}
+2\frac{M_{p}^2V_{TT}}{V^2} \nonumber \\
&=& -\frac{2}{s}\frac{1+3x^2}{x(1+x^2)^{5/2}}\,,
\label{nSeq}
\end{eqnarray}
whereas the ratio of tensor to scalar perturbations is
\begin{eqnarray}
r=8\frac{V_{T}^2M_{p}^2}{V^3}
=\frac{8}{s}\frac{1}{x(x^2+1)^{5/2}}\,.
\label{ratio}
\end{eqnarray}
%

\begin{figure}
\includegraphics[height=2.8in,width=3.4in]{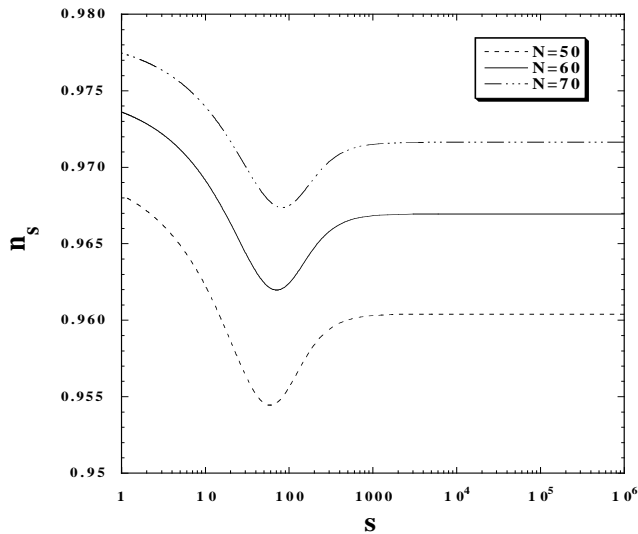}
\caption{The spectral index $n_{\rm S}$ of scalar
metric perturbations as a function of $s$ with three
different number of $e$-foldings ($N=50, 60, 70$).
This figure corresponds to the case in which inflation
ends in the region $x_{f} \ll 1$.
}
\label{ns}
\end{figure}

\begin{figure}
\includegraphics[height=2.8in,width=3.4in]{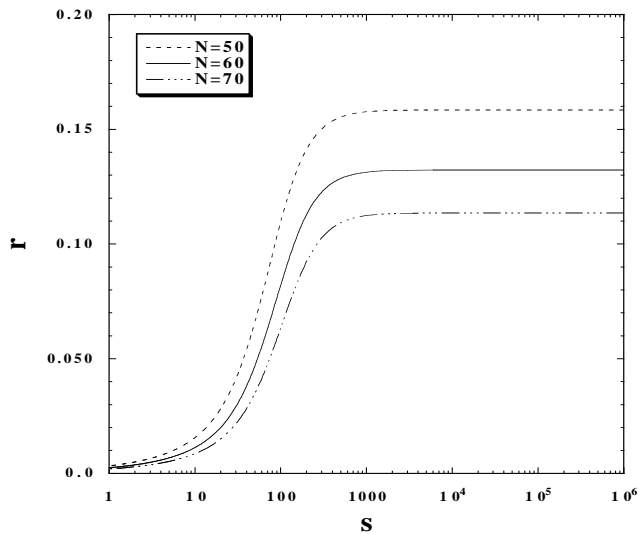}
\caption{The tensor-to-scalar ratio $r$ as a function
of $s$ with three different number
of $e$-foldings ($N=50, 60, 70$).
}
\label{ratiofig}
\end{figure}

We shall study the case in which the end of inflation corresponds
to the region with an exponential potential, i.e., $x_{f} \ll 1$.
When $s=1$, Eq.~(\ref{ep}) shows that inflation
ends around $x_{f} \sim 0.5$.
Hence the approximation, $x_{f} \ll 1$, is valid when
$s$ is larger than of order unity.
In this case one has $x_{f} \simeq 1/2s$ from Eq.~(\ref{ep}).
Since $f(x) \simeq x$ for $x \ll 1$, we find
\begin{eqnarray}
\label{fNs}
f(x)=(N+1/2)/s\,.
\end{eqnarray}

In the regime of an exponential potential ($x \ll 1$)
we have $sx \simeq N+1/2$.
In this case Eqs.~(\ref{nSeq}) and (\ref{ratio}) give
\begin{eqnarray}
\label{nseq}
n_{\rm S}-1 &=& -\frac{4}{2N+1}\,, \\
\label{ratioeq}
r&=&\frac{16}{2N+1}\,.
\end{eqnarray}
Hence $n_{\rm S}$ and $r$ are
dependent on the number of $e$-foldings only.
{}From Eqs.~(\ref{nseq}) and (\ref{ratioeq})
we find that $n_{\rm S}=0.9669$ and
$r=0.1322$ for $N=60$. It was shown in Ref.~\cite{GST} that this
case is well inside the $1\sigma$ contour bound
coming from the observational constraints of
WMAP, SDSS and 2dF (see also Ref.~\cite{SV04}).

\begin{figure}
\includegraphics[height=2.8in,width=3.4in]{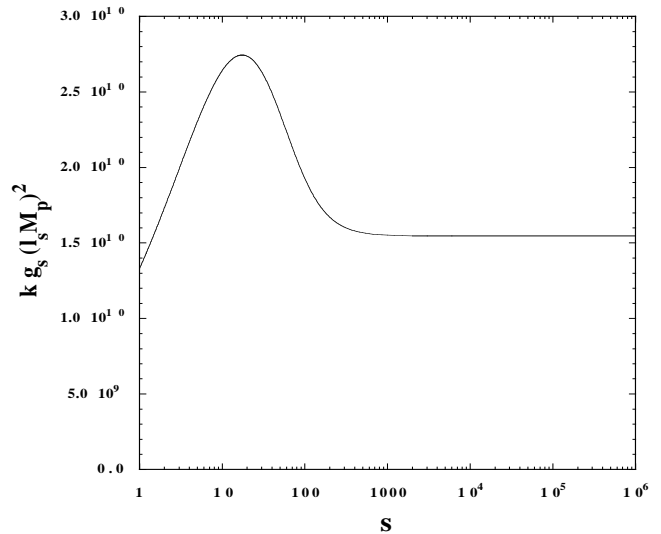}
\caption{The quantity $kg_{s} (l_{s} M_{p})^2$
as a function of $s$. This is derived by the COBE
normalization at $N=60$.}
\label{kgs}
\end{figure}

Of course there is a situation in which cosmologically
relevant scales ($55 \lesssim N \lesssim 65$)
correspond to the region $x \gtrsim 1$.
In Figs.~\ref{ns} and \ref{ratiofig} we plot $n_{\rm S}$
and $r$ as a function of $s$ for three different values of
$N$. For large $s (\gg 1)$, we find that
the quantity $x$ is much smaller
than unity from the relation (\ref{fNs}).
Hence $n_{\rm S}$ and $r$ are given by the
formula (\ref{nseq}) and (\ref{ratioeq}).
For smaller $s$ the quantity $x$ becomes
larger than of order unity,
which means that the results (\ref{nseq}) and
(\ref{ratioeq}) can no longer be used.
In Fig.~\ref{ns} we find that the spectral index has
a minimum around $s=70$ for $N=60$.
This roughly corresponds to the region $x=R/L \sim 1$.
As we see from Fig.~\ref{potential} the potential becomes
flatter for $x \gtrsim 1$. This leads to the increase of
the spectral index toward $n_{\rm S}=1$
with the decrease of $s$.
Recent observations show that
$n_{\rm S}=0.98 \pm 0.02$ at the 95\% confidence
level \cite{Seljak} (see also Refs.~\cite{inobser}).
As we find in Fig.~\ref{ns} this condition is satisfied
for $N \gtrsim 60$.

The tensor-to-scalar ratio is given by Eq.~(\ref{ratioeq})
for $s \gg 1$. For a fixed value of $N$
this ratio gets smaller with the decrease of
$s$. This is understandable, since the potential becomes flatter
as we enter the region $x \gtrsim 1$.
The tensor-to-scalar ratio is constrained to be $r<0.36$
at the 95\% confidence level from recent observations \cite{Seljak}.
Hence our model satisfies this observational constraint.

When $x_{60} \ll 1$ the condition of the COBE
normalization (\ref{COBE}) gives
\begin{eqnarray}
\label{nseq2}
kg_s(l_sM_{p})^2 \simeq \frac{10^9}{24\pi^2}
(60+1/2)^2 \simeq 1.55 \times 10^{10}\,,
\end{eqnarray}
which is independent of $s$.
As we see from Fig.~\ref{kgs} the quantity $kg_s(l_sM_{p})^2$
departs from the value (\ref{nseq2}) for smaller $s$.
However $kg_s(l_sM_{p})^2$ is of order $10^{10}$
for $s \gtrsim 1$.
It is interesting to note that the COBE normalization
uniquely fixes the value of the
potential at the end of inflation if it happens in the
regime of an exponential potential
independently of the fact where inflation had commenced.
In fact using Eq.~(\ref{pox}) gives
\begin{eqnarray}
\label{vend}
V_{\rm end} \simeq x_{f}\tau_{3}=
\frac{1}{2kg_s (l_s M_p)^2}
M_p^4 \simeq 3.2 \times 10^{-11}M_p^4\,.
\end{eqnarray}
This sets the energy scale to be $ V^{1/4}_{\rm end} \simeq 2.3 \times
10^{-3} M_p$.

The above discussion corresponds to the case in which
inflation ends in the region $x_{f} \ll 1$.
In order to understand the behavior of another asymptotical
region, let us consider a situation
when inflation ends for $x_{f} \gg 1$.
In this case the end of inflation is characterized by
$x_f^6 \simeq 1/(2s)$. Since $x_{f} \gg 1$, we are
considering a parameter range $s \ll 1$.
When $x \gg 1$ the function $f(x)$ behaves as
$f(x) \simeq x^4/4$, which gives the relation
$x^4 \simeq 4N/s$.
Hence we obtain
\begin{eqnarray}
\label{nslarge}
n_{\rm S}-1 &=&-\frac{3}{2N}\,, \\
r &=& \frac{\sqrt{s}}{N^{3/2}}\,, \\
kg_s(l_sM_{p})^2 &=&
\frac{10^9 \sqrt{2} N^{3/2}}{6\pi^2}
\sqrt{s}\,.
\end{eqnarray}

While $n_{\rm S}$ is independent of $s$,
both $r$ and $kg_s(l_sM_{p})^2$ are dependent
on $s$ and $N$.
For example one has $n_{\rm S}=0.975$,
$r=0.003\sqrt{s}$ and
$kg_s(l_sM_{p})^2=1.11 \times 10^{10} \sqrt{s}$
for $N=60$.
{}From Fig.~\ref{ns} we find that $n_{\rm S}$ increases
with the decrease of $s$ in the region
$1 \lesssim s \lesssim 50$ for a fixed $N$.
This tendency persists for $s \lesssim 1$ and
$n_{\rm S}$ approaches a constant value
given by Eq.~(\ref{nslarge}) as $s$ decreases.
We note that the spectral index $n_{\rm S}$
satisfies the observational constraint coming from
recent observations.
The tensor-to-scalar ratio is strongly suppressed
in the region $s < 1$, which also satisfies
the observational constraint.
The quantity $kg_s(l_sM_{p})^2$ gets smaller with
the decrease of $s$.

We can estimate the
the potential energy at the end of inflation
in the regime described by $x_f \gg 1$, as
\begin{eqnarray}
V_{\rm end} &\simeq& \tau_3 \,, \\
\tau_3 &=& \frac{s}{kg_{s} (l_{s}M_{p})^2}M_p^4
\simeq 9.0 \times 10^{-11} \sqrt{s} M_p^4\,.
\end{eqnarray}
In this case $V_{\rm end}$ depends on the value of $s$.
The order of the energy scale does not differ from
(\ref{vend}) provided that $s$ is not too much smaller
than unity.

In summary we find that $n_{\rm S}$ and $r$ in our model
satisfy observational constraints of CMB for any values of $s$,
which means that we do not obtain the constraint on $s$.
This is different from the geometrical
tachyon inflation with potential $V=V_{0} \cos (T/\sqrt{kl_s^2})$
in which the spectral index $n_{\rm S}$
provides constrains on model parameters \cite{PST}.
The only constraint in our model is the COBE normalization.
If we demand that the value of $R$ at the end of inflation
is larger than $l_{s}$, this gives
\begin{eqnarray}
\label{nseq3}
k<16\pi^6 g_s \left(\frac{M_{p}}{M_{s}}\right)^4\,,
\end{eqnarray}
where we used $\tau_{3}=M_{s}^4/(2\pi)^3 g_s$.

Combining this relation with the condition of the
COBE normalization:
$kg_s(l_sM_{p})^2 \simeq 10^{10}$ for $s \gtrsim 1$,
we find
\begin{eqnarray}
g_{s}> \frac{10^5}{4\pi^3}
\left(\frac{M_{s}}{M_{p}}\right)^3\,.
\end{eqnarray}
Since we require the condition $g_{s} \ll 1$ for the validity
of the theory, this gives the constraint
\begin{eqnarray}
\label{mcon}
M_s/M_p \ll 0.1\,.
\end{eqnarray}

After the field reaches the point $R=l_s$, we assume that
the field $T$ is frozen at this point, which is a reasonable
assumption given what we understand from the bulk description
of the dynamics.
This gives us a positive cosmological constant in the system.

\section{After the end of inflation}

The first phase driven by the field $T$ is triggered by the
second phase driven by the field $\chi$.
Introducing new variables $\chi=\chi_1+i \chi_2$,
$X^2=\chi_1^2+\chi_2^2$,
$\tilde{X}=M_p X$ and $\tilde{v}=M_p^2v$,
the potential (\ref{chipot}) of the field $X$ reduces to
\begin{equation}
U(\tX) =  \frac{1}{4\pi^4 (l_sM_{p})^4 g_s k}
\left[ \pi (k+1) \tX^4-\tv \tX^2 \right]\,.
\end{equation}
This potential has two local minima at
$\tX_{c}=\pm \sqrt{\tv/(2\pi (k+1))}$
with negative energy
\begin{equation}
U(\tX_{c}) =
-\frac{\tv^2}{16\pi^5 k(k+1)(l_{s}M_{p})^4g_{s}}\,.
\end{equation}
One can cancel (or nearly cancel) this term by taking into
account the energy of the field $T$ at $R=l_s$.
Since this is given by $V(R=l_s)=\tau_3/\sqrt{kg_s}$,
the condition $V(R=l_s)+U(\tX_{c})=0$ leads to
\begin{eqnarray}
\tv^2 =
16\pi^5 (k+1)\tau_{3}(l_sM_p)^4\sqrt{kg_s}\,.
\end{eqnarray}
Using the relation $\tau_{3}=M_s^4/(2\pi)^3 g_s$,
this can be written as
\begin{eqnarray}
\label{vnew}
{\tv}^2=\frac{2\pi^2 \sqrt{k} (k+1)}{\sqrt{g_s}}\,.
\end{eqnarray}

Then the total potential of our system is
\begin{equation}
\label{dwell}
W=A \left(\tX^2-\tX_c^2\right)^2\,,
\end{equation}
where
\begin{equation}
A \equiv \frac{k+1}{4\pi^3 (l_{s} M_{p})^4 g_s k}\,.
\end{equation}
The mass of the potential at $\tX=0$ is given by
\begin{equation}
\label{nmass}
m^2 \equiv
\frac{{\rm d}^2 W}{{\rm d} \tX^2} (\tX=0)
=-4A\tX_c^2\,.
\end{equation}
Meanwhile the square of the Hubble constant at $\tX=0$ is
\begin{equation}
H_0^2=\frac{A\tv^2}{12\pi^2 (k+1)^2M_p^2}\,.
\end{equation}
Then we obtain the following ratio
\begin{equation}
\label{massratio}
\frac{|m^2|}{H_0^2}=\frac{24\pi (k+1)}{v}
=\frac{12\sqrt{2}(k+1)^{1/2}}{k^{1/4}}g_s^{1/4}\,,
\end{equation}
where we used Eq.~(\ref{vnew}) in the second equality.

As we showed in the previous section, the COBE normalization
gives $kg_s(l_s M_p)^2 \simeq 10^{10}$ for $s \gtrsim 1$.
Then the ratio (\ref{massratio}) can be estimated as
\begin{eqnarray}
\frac{|m^2|}{H_0^2} &\simeq&
5 \times 10^3 \left(\frac{k+1}{k}\right)^{1/2}
\left(\frac{M_s}{M_p}\right)^{1/2} \nonumber \\
&\simeq& 5 \times 10^3
\left(\frac{M_s}{M_p}\right)^{1/2}\,.
\end{eqnarray}
Then we have $|m^2|>H_0^2$ for
\begin{equation}
\label{stmass}
M_s/M_p>4 \times 10^{-8}\,.
\end{equation}
This means that the second stage of inflation does not occur for the field
$\chi$ provided that the string mass scale $M_s$ satisfies the
condition (\ref{stmass}).
When $4 \times 10^{-8}<M_s/M_p \ll 10^{-1}$, inflation ends
before the field $T$ reaches the point $R=l_{s}$,
which is triggered by a fast roll of the field $\chi$.
This situation is similar to the original hybrid inflation
model \cite{Linde94}.

When $M_s/M_p<4 \times 10^{-8}$, double inflation
occurs even after the end of the first stage of
inflation. In this case the CMB constraints discussed
in the previous section need to be modified.
However the second stage of inflation
is absent for the natural string mass scale which is not too much
smaller than the Planck scale.

We note that the vacuum expectation value of the field
$\tX$ is given by
\begin{equation}
{\tX}_c=2\sqrt{3} \frac{H_0}{|m|}M_{p}\,.
\end{equation}
When $|m| \gtrsim H_0$ we find that $\tX_c$ is less than
of order the Planck mass.
When double inflation occurs ($|m| \lesssim H_0$),
the amplitude of symmetry breaking
takes a super-Planckian value $\tX_{c} \gtrsim M_{p}$.
In this sense the latter case does not look natural compared
to the case in which the second stage of inflation does not occur.

Since the field $\chi$ has a standard kinematic term, reheating
proceeds as in the case of potentials with spontaneous
symmetry breaking. This is in contrast to a tachyon field
governed by the DBI action in which the energy density of the tachyon
overdominates the universe soon after the end of inflation.
Thus the problem of reheating present in DBI tachyon
models \cite{KL02,Frolov} is absent in our model.
Since the potential of the field $X$ has a negative mass
given by Eq.~(\ref{nmass}), this leads to the exponential growth of quantum
fluctuations of $X$ with momenta $k<|m|$, i.e.,
$\delta X_{k} \propto
\exp(\sqrt{|m^2|-k^2}\,t)$ \cite{Boy}.
This negative instability is so strong that one can not trust
perturbation theory including the Hartree and $1/N$ approximations.
We require lattice simulations in order to take into account
rescattering of created particles and the production
of topological defects \cite{Felder}.

It was shown in Refs.~\cite{Felder} that symmetry breaking
ends after one oscillation of the field distribution as the field
evolves toward the potential minimum.
This refects the fact that gradient energies of
all momentum modes do not return back to
the original state at $X=0$ because of a very complicated
field distribution after the violent growth of
quantum fluctuations.

Finally we should mention that de-Sitter vacua can be obtained
provided that the potential energy
$V(R=l_s)$ does not exactly cancel
the negative energy $U(\tX_{c})$.
In order to match with the current energy scale of dark energy, we
require an extreme fine tuning
$V(R=l_s)+U(\tX_{c}) \simeq 10^{-123}M_p^4$.
However this kind of fine tuning is a generic
problem of dark energy.

\section{Conclusions}

In this paper we studied the motion of a BPS D3-brane
in the presence of a stack of $k$ parallel D5-branes
in type II string theory.
Inflation is realized by the potential energy of a radion
field $R$ which characterizes the distance of
D3 and D5 branes.
This potential is not in general written explicitly, but
is approximately given by
(\ref{pot2}) for $R \ll \sqrt{kg_s}l_s$
and (\ref{pot}) for $R \gg \sqrt{kg_s}l_s$.
We evaluated the spectral index of scalar metric perturbations
and the tensor-to-scalar ratio together
with the number of $e$-foldings under the condition
that inflation ends in the region $R \ll \sqrt{kg_s}l_s$.
This model satisfies observational constraints
coming from CMB, SDSS and 2dF
independently of the value of $s$ defined by Eq.~(\ref{sdef}).
We also note that this result does not change even
when inflation ends in the region $R \gg \sqrt{kg_s}l_s$.

The only strong constraint coming from CMB is the
COBE normalization, i.e., $kg_{s}(l_{s}M_{p})^2
\simeq 10^{10}$ for $s \gtrsim 1$.
If we demand that the inflationary period is over
before the radion reaches the point $R=l_s$,
this gives the constraint on the number of
D5-branes; see Eq.~(\ref{nseq3}).
Combining this with the condition of the COBE normalization,
the string mass scale is constrained to be $M_s/M_p \ll 0.1$
for the validity of the weak-coupling approximation ($g_s \ll 1$).

When the radion field enters the region $R \lesssim l_s$,
the description of closed string background is no longer valid.
Instead the dynamics should be studied using
a complex scalar field $\chi$
living on the world volume of open strings
stretched between the probe D3-brane and the D5-branes.
We assumed that the radion field is frozen in the region
$R \lesssim  l_{s}$, which gives rise to
a positive cosmological constant.
The potential of the field $\chi$ is given by Eq.~(\ref{chipot}),
which has a negative energy at the potential minimum.
If this energy is cancelled by the positive cosmological constant,
we obtain the double-well potential given by Eq.~(\ref{dwell}).

We found that the absolute value of the mass of this double-well
potential at $\chi=0$ is larger than the Hubble parameter
provided that $M_s/M_p>4 \times 10^{-8}$.
Hence in this case the second stage of inflation does not occur
and the evolution of the field $\chi$ is described by a fast roll.
Since the action of the complex field has a standard kinematic term,
the problem of reheating present in DBI tachyon models is
absent in our scenario.
Reheating in our model is described by tachyonic preheating
in which quantum fluctuations grow exponentially
by a negative instability.
The symmetry breaking would
end after one oscillation of the field distribution as the field
evolves toward the potential minimum.

It is also possible to explain the origin of dark energy
if a positive cosmological constant does not exactly
cancel the negative potential energy of the field $\chi$.
Although this requires a fine tuning, it is intriguing that
our model provides a number of promising ways to
provide viable cosmological evolution.

One of the potential problems with our model is
that the compactification is not necessarily realistic.
Although we can encode the physics of the gravity
background as a non-trivial scalar field on a flat brane, we are treating
this latter object as being fundamental.
Thus by compactifying this on a $T^6$ we will
be missing higher order terms coming from the full compactification of the
D5-solution. These terms may play a more important
role in the cosmological theory on the D3-brane.
It may be useful to compare the results obtained in this paper with a full
string compactification by smearing the SUGRA
harmonic function on a $T^4$ and compactifying the remaining directions on
the two-cycles of a torus. The resultant analysis is complicated since the
DBI may not be valid, however this is beyond the scope of the current
endeavor.

\section*{ACKNOWLEDGMENTS}
We thank Ashoke Sen for useful discussions. S.\,T. is supported by
JSPS (Grant No.\,30318802). J.\,W. thanks S.\,Thomas and
is supported by a QMUL studentship.
S.\,P. thanks Y.\,Kitazawa and the Theory Group, KEK,
Tsukuba for a visiting fellowship and warm hospitality.
This work has been carried out during this period.



\begin{thebibliography}{99}

\bibitem{senrev}
A.~Sen,
arXiv:hep-th/0410103.

\bibitem{DBI}
A.~Sen, JHEP {\bf 9910}, 008 (1999); M.~R.~Garousi, Nucl. Phys.
B{\bf 584}, 284 (2000); Nucl. Phys. B {\bf 647}, 117 (2002); JHEP
{\bf 0305}, 058 (2003); E.~A.~Bergshoeff, M.~de Roo, T.~C. de Wit,
E.~Eyras, S.~Panda, JHEP {\bf 0005}, 009 (2000); J.~Kluson, Phys.
Rev. D {\bf 62}, 126003 (2000); D.~Kutasov and V.~Niarchos, Nucl.
Phys. B {\bf 666}, 56 (2003).

\bibitem{kutasov}
D.~Kutasov, arXiv:hep-th/0405058;
arXiv:hep-th/0408073.

\bibitem{Sahakyan}
D.~A.~Sahakyan,
JHEP {\bf 0410}, 008 (2004).

\bibitem{pani}
K.~L.~Panigrahi,
Phys.\ Lett.\ B {\bf 601}, 64 (2004).

\bibitem{TW}
S.~Thomas and J.~Ward,
JHEP {\bf 0502}, 015 (2005);
S.~Thomas and J.~Ward,
JHEP {\bf 0510}, 098 (2005).

\bibitem{tachinfl}
A.~Mazumdar, S.~Panda and A.~Perez-Lorenzana,
Nucl.\ Phys.\ B {\bf 614}, 101 (2001);
M.~Fairbairn and M.~H.~G.~Tytgat,
Phys.\ Lett.\ B {\bf 546}, 1 (2002); A.~Feinstein,
Phys.\ Rev.\ D {\bf 66}, 063511 (2002);
M.~Sami, P.~Chingangbam and T.~Qureshi,
Phys.\ Rev.\ D {\bf 66}, 043530 (2002); M.~Sami,
Mod.\ Phys.\ Lett.\ A {\bf 18}, 691 (2003); Y.~S.~Piao, R.~G.~Cai,
X.~m.~Zhang and Y.~Z.~Zhang,
Phys.\ Rev.\ D {\bf 66}, 121301 (2002).

\bibitem{inflation}
A.~Linde, {\em Particle Physics and Inflationary Cosmology},
Harwood, Chur (1990)
[arXiv:hep-th/0503203];
D.~H.~Lyth and A.~Riotto,
Phys.\ Rept.\  {\bf 314}, 1 (1999);
A.~R.~Liddle and D.~H. ~Lyth,
{\em Cosmological inflation and large-scale structure},
Cambridge University Press (2000);
B.~A.~Bassett, S.~Tsujikawa and D.~Wands,
arXiv:astro-ph/0507632.

\bibitem{WMAP}
D.~N.~Spergel {\it et al.},
Astrophys.\ J.\ Suppl.\  {\bf 148}, 175 (2003);
H.~V.~Peiris \textit{et al.}, Astrophys. J. Suppl.
\textbf{148}, 213 (2003).

\bibitem{SDSS}
M.~Tegmark {\it et al.},
Phys.\ Rev.\ D {\bf 69}, 103501 (2004);
K.~Abazajian {\it et al.},
Astron.\ J.\  {\bf 128}, 502 (2004).

\bibitem{2dF}
W.~J.~Percival {\it et al.},
Mon.\ Not.\ Roy.\ Astron.\ Soc.\  {\bf 327}, 1297 (2001);
S.~Cole {\it et al.},
Mon.\ Not.\ Roy.\ Astron.\ Soc.\  {\bf 362}, 505 (2005).

\bibitem{general}
F.~Quevedo,
Class. Quant. Grav. {\bf 19}, 5721-5779 (2002);
A.~Linde,
J. Phys. Conf. Ser. {\bf 24}, 151-160 (2005).

\bibitem{SNIa}
S.~Perlmutter {\it et al.},
Astrophys.\ J.\  {\bf 517}, 565 (1999);
A.~G.~Riess {\it et al.},
Astron.\ J.\  {\bf 116}, 1009 (1998);
Astron.\ J.\  {\bf 117}, 707 (1999).

\bibitem{darkenergy}
R.~Bousso and J.~Polchinski,
JHEP {\bf 0006}, 006 (2000);
T.~Banks and M.~Dine,
JHEP {\bf 0110}, 012 (2001);
A.~Albrecht, C.~P.~Burgess, F.~Ravndal and C.~Skordis,
  Phys.\ Rev.\ D {\bf 65}, 123507 (2002);
S.~Kachru, R.~Kallosh, A.~Linde and S.~P.~Trivedi,
Phys.\ Rev.\ D {\bf 68}, 046005 (2003);
R.~Kallosh and A.~Linde,
Phys.\ Rev.\ D {\bf 67}, 023510 (2003);
C.~P.~Burgess, R.~Kallosh and F.~Quevedo,
JHEP {\bf 0310}, 056 (2003);
C.~P.~Burgess,
AIP Conf.\ Proc.\  {\bf 743}, 417 (2005)
[arXiv:hep-th/0411140];
I.~Ya.~Aref'eva,
arxiv:astro-ph/0410443;
E.~J.~Copeland, M.~Sami and S.~Tsujikawa,
arXiv:hep-th/0603057.

\bibitem{cospep}
G.~W.~Gibbons,
Phys.\ Lett.\ B {\bf 537}, 1 (2002);
S.~Mukohyama,
Phys.\ Rev.\ D {\bf 66}, 024009 (2002); Phys.\ Rev.\ D {\bf 66},
123512 (2002);
D.~Choudhury, D.~Ghoshal, D.~P.~Jatkar and S.~Panda,
Phys.\ Lett.\ B {\bf 544}, 231 (2002);
G.~Shiu and I.~Wasserman,
Phys.\ Lett.\ B {\bf 541}, 6 (2002);
T.~Padmanabhan,
Phys.\ Rev.\ D {\bf 66}, 021301 (2002);
J.~S.~Bagla, H.~K.~Jassal and T.~Padmanabhan,
Phys.\ Rev.\ D {\bf 67}, 063504 (2003);
G.~N.~Felder, L.~Kofman and A.~Starobinsky,
JHEP {\bf 0209}, 026 (2002); J.~M.~Cline, H.~Firouzjahi and
P.~Martineau,
JHEP {\bf 0211}, 041 (2002);
J.~g.~Hao and X.~z.~Li,
Phys.\ Rev.\ D {\bf 66}, 087301 (2002);
C.~j.~Kim, H.~B.~Kim and Y.~b.~Kim,
Phys.\ Lett.\ B {\bf 552}, 111 (2003);
T.~Matsuda,
Phys.\ Rev.\ D {\bf 67}, 083519 (2003);
A.~Das and A.~DeBenedictis,
arXiv:gr-qc/0304017;
Z.~K.~Guo, Y.~S.~Piao, R.~G.~Cai and Y.~Z.~Zhang,
Phys.\ Rev.\ D {\bf 68}, 043508 (2003);
L.~R.~W.~Abramo and F.~Finelli,
Phys.\ Lett.\ B {\bf 575} (2003) 165;
G.~W.~Gibbons,
Class.\ Quant.\ Grav.\  {\bf 20}, S321 (2003);
M.~Majumdar and A.~C.~Davis,
Phys.\ Rev.\ D {\bf 69}, 103504 (2004);
S.~Nojiri and S.~D.~Odintsov,
Phys.\ Lett.\ B {\bf 571}, 1 (2003);
E.~Elizalde, J.~E.~Lidsey, S.~Nojiri and S.~D.~Odintsov,
Phys.\ Lett.\ B {\bf 574}, 1 (2003);
V.~Gorini, A.~Y.~Kamenshchik, U.~Moschella and V.~Pasquier,
Phys.\ Rev.\ D {\bf 69}, 123512 (2004);
L.~P.~Chimento,
Phys.\ Rev.\ D {\bf 69}, 123517 (2004);
J.~M.~Aguirregabiria and R.~Lazkoz,
Phys.\ Rev.\ D {\bf 69}, 123502 (2004);
M.~B.~Causse,
arXiv:astro-ph/0312206;
B.~C.~Paul and M.~Sami,
Phys.\ Rev.\ D {\bf 70}, 027301 (2004);
G.~N.~Felder and L.~Kofman,
Phys.\ Rev.\ D {\bf 70}, 046004 (2004);
J.~M.~Aguirregabiria and R.~Lazkoz,
Mod.\ Phys.\ Lett.\ A {\bf 19}, 927 (2004);
L.~R.~Abramo, F.~Finelli and T.~S.~Pereira,
Phys.\ Rev.\ D {\bf 70}, 063517 (2004);
G.~Calcagni,
Phys.\ Rev.\ D {\bf 70}, 103525 (2004);
G.~Calcagni and S.~Tsujikawa,
Phys.\ Rev.\ D {\bf 70}, 103514 (2004);
J. Raemaekers, JHEP {\bf 0410}, 057 (2004);
P.~F.~Gonzalez-Diaz,
Phys.\ Rev.\ D {\bf 70}, 063530 (2004);
S.~K.~Srivastava,
arXiv:gr-qc/0409074; gr-qc/0411088;
P.~Chingangbam and T.~Qureshi,
Int.\ J.\ Mod.\ Phys.\ A {\bf 20}, 6083 (2005);
S.~Tsujikawa and M.~Sami,
Phys.\ Lett.\ B {\bf 603}, 113 (2004);
M.~R.~Garousi, M.~Sami and
S.~Tsujikawa, Phys.\ Lett.\ B {\bf 606}, 1 (2005);
Phys.\ Rev.\ D {\bf 71}, 083005 (2005);
N.~Barnaby and J.~M.~Cline,
Int.\ J.\ Mod.\ Phys.\ A {\bf 19}, 5455 (2004); E.~J.~Copeland,
M.~R.~Garousi, M.~Sami and S.~Tsujikawa,
Phys.\ Rev.\ D {\bf 71}, 043003 (2005);
B.~Gumjudpai, T.~Naskar, M.~Sami and S.~Tsujikawa,
JCAP {\bf 0506}, 007 (2005);
M.~Novello, M.~Makler, L.~S.~Werneck and C.~A.~Romero,
Phys.\ Rev.\ D {\bf 71}, 043515 (2005); A.~Das, S.~Gupta,
T.~D.~Saini and S.~Kar,
Phys.\ Rev.\ D {\bf 72}, 043528 (2005);
H.~Singh,
arXiv:hep-th/0505012;
S.~Tsujikawa,
arXiv:astro-ph/0508542;
P.~Chingangbam, S.~Panda and A.~Deshamukhya,
JHEP {\bf 0502}, 052 (2005);
D.~Cremades, F.~Quevedo and A.~Sinha,
JHEP {\bf 0510}, 106 (2005);
L.~Amendola, S.~Tsujikawa and M.~Sami,
Phys.\ Lett.\ B {\bf 632}, 155 (2006);
G.~Calcagni,
arXiv:hep-th/0512259;
A.~Ghodsi and A.~E.~Mosaffa,
Nucl.\ Phys.\ B {\bf 714}, 30 (2005).

\bibitem{KL02}
L.~Kofman and A.~Linde,
JHEP {\bf 0207}, 004 (2002).

\bibitem{Frolov}
A.~V.~Frolov, L.~Kofman and A.~A.~Starobinsky,
Phys.\ Lett.\ B {\bf 545}, 8 (2002).

\bibitem{Dvali}
G.~R.~Dvali and S.~H.~H.~Tye,
Phys.\ Lett.\ B {\bf 450}, 72 (1999);
G.~R.~Dvali, Q.~Shafi and S.~Solganik,
arXiv:hep-th/0105203.

\bibitem{braneantibrane}
C.~P.~Burgess, M.~Majumdar, D.~Nolte,
F.~Quevedo, G.~Rajesh and R.~J.~Zhang,
JHEP {\bf 0107}, 047 (2001);
C.~P.~Burgess, P.~Martineau, F.~Quevedo, G.~Rajesh and R.~J.~Zhang,
JHEP {\bf 0203}, 052 (2002);
D.~Choudhury, D.~Ghoshal, D.~P.~Jatkar
and S.~Panda,
JCAP {\bf 0307}, 009 (2003).

\bibitem{Garcia}
J.~Garcia-Bellido, R.~Rabadan and F.~Zamora,
JHEP {\bf 0201}, 036 (2002);
N.~Jones, H.~Stoica and S.~H.~H.~Tye,
JHEP {\bf 0207}, 051 (2002);
M.~Gomez-Reino and I.~Zavala,
JHEP {\bf 0209}, 020 (2002).

\bibitem{Hirano}
C.~Herdeiro, S.~Hirano and R.~Kallosh,
JHEP {\bf 0112}, 027 (2001).

\bibitem{kklmmt}
S.~Kachru, R.~Kallosh, A.~Linde, J.~Maldacena,
L.~McAllister and S.~P.~Trivedi,
JCAP {\bf 0310}, 013 (2003).

\bibitem{dbranepapers}
C.~P.~Burgess, J.~M.~Cline, H.~Stoica and F.~Quevedo,
JHEP {\bf 0409}, 033 (2004);
O.~DeWolfe, S.~Kachru and H.~L.~Verlinde,
JHEP {\bf 0405}, 017 (2004);
N.~Iizuka and S.~P.~Trivedi,
Phys.\ Rev.\ D {\bf 70}, 043519 (2004);
A.~Buchel and A.~Ghodsi,
Phys.\ Rev.\ D {\bf 70}, 126008 (2004);
J.~J.~Blanco-Pillado {\it et al.},
JHEP {\bf 0411}, 063 (2004);
M.~Berg, M.~Haack and B.~Kors,
Phys.\ Rev.\ D {\bf 71}, 026005 (2005); 
M.~Berg, M.~Haack and B.~Kors, hep-th/0404087.

\bibitem{D3D7}
K.~Dasgupta, C.~Herdeiro, S.~Hirano and R.~Kallosh,
Phys.\ Rev.\ D {\bf 65}, 126002 (2002);
J.~P.~Hsu, R.~Kallosh and S.~Prokushkin,
JCAP {\bf 0312}, 009 (2003);
F.~Koyama, Y.~Tachikawa and T.~Watari,
Phys.\ Rev.\ D {\bf 69}, 106001 (2004)
[Erratum-ibid.\ D {\bf 70}, 129907 (2004)];
K.~Dasgupta, J.~P.~Hsu, R.~Kallosh, A.~Linde and M.~Zagermann,
JHEP {\bf 0408}, 030 (2004);
H.~Firouzjahi and S.~H.~H.~Tye,
Phys.\ Lett.\ B {\bf 584}, 147 (2004).

\bibitem{TWinf}
S.~Thomas and J.~Ward,
Phys.\ Rev.\ D {\bf 72}, 083519 (2005).

\bibitem{PST}
S.~Panda, M.~Sami and S.~Tsujikawa,
arXiv:hep-th/0510112.

\bibitem{mirage}
A.~Kehagias and E.~Kiritsis,
JHEP {\bf 9911}, 022 (1999).

\bibitem{branonium}
C.~P.~Burgess, P.~Martineau, F.~Quevedo and R.~Rabadan,
JHEP {\bf 0306}, 037 (2003);
C.~P.~Burgess, N.~E.~Grandi, F.~Quevedo and R.~Rabadan,
JHEP {\bf 0401}, 067 (2004);
K.~Takahashi and K.~Ichikawa,
Phys. \ Rev. \ D {\bf 69}, 103506 (2004).

\bibitem{Callan}
C.~G.~Callan, J.~A.~Harvey and A.~Strominger,
Nucl.\ Phys.\ B {\bf 367}, 60 (1991).

\bibitem{sen2}
A.~Sen,
Mod.\ Phys.\ Lett.\ A {\bf 17}, 1797 (2002).

\bibitem{Gava}
E.~Gava, K.~S.~Narain and M.~H.~Sarmadi,
Nucl.\ Phys.\ B {\bf 504}, 214 (1997).

\bibitem{tong}
E.~Silverstein and D.~Tong,
Phys.\ Rev.\ D {\bf 70}, 103505 (2004).


\bibitem{Hwang}
J.~c.~Hwang and H.~Noh,
Phys.\ Rev.\ D {\bf 66}, 084009 (2002).

\bibitem{SV04}
D.~A.~Steer and F.~Vernizzi,
Phys.\ Rev.\ D {\bf 70}, 043527 (2004).

\bibitem{GST}
M.~R.~Garousi, M.~Sami and S.~Tsujikawa,
Phys.\ Rev.\ D {\bf 70}, 043536 (2004).

\bibitem{Seljak}
U.~Seljak {\it et al.},
Phys.\ Rev.\ D {\bf 71}, 103515 (2005).

\bibitem{inobser}
V.~Barger, H.~S.~Lee, and D.~Marfatia,
Phys.\ Lett. \ B \textbf{565}, 33 (2003);
W.~H.~Kinney, E.~W.~Kolb, A.~Melchiorri and A.~Riotto,
Phys.\ Rev.\ D \textbf{69}, 103516 (2004);
S.~M.~Leach and A.~R.~Liddle, Phys. Rev. D
\textbf{68}, 123508 (2003);
M. Tegmark {\em et al.},
Phys.\ Rev.\ D {\bf 69}, 103501 (2004);
S.~Tsujikawa and A.~R.~Liddle,
JCAP {\bf 0403}, 001 (2004);
S.~Tsujikawa and B.~Gumjudpai,
Phys.\ Rev.\ D {\bf 69}, 123523 (2004);
L.~Alabidi and D.~H.~Lyth,
arXiv:astro-ph/0510441;
arXiv:astro-ph/0603539.

\bibitem{Linde94}
A.~D.~Linde,
Phys.\ Rev.\ D {\bf 49}, 748 (1994).

\bibitem{Boy}
D.~Boyanovsky {\it et al.},
Phys.\ Rev.\ D {\bf 57}, 2166 (1998);
D.~Cormier and R.~Holman,
Phys.\ Rev.\ D {\bf 60}, 041301 (1999);
S.~Tsujikawa and T.~Torii,
Phys.\ Rev.\ D {\bf 62}, 043505 (2000).

\bibitem{Felder}
G.~N.~Felder {\it et al.},
Phys.\ Rev.\ Lett.\  {\bf 87}, 011601 (2001);
G.~N.~Felder, L.~Kofman and A.~D.~Linde,
Phys.\ Rev.\ D {\bf 64}, 123517 (2001).


\end{thebibliography}
\end{document}